\documentclass[aps,reprint,superscriptaddress,prx]{revtex4-1}
\usepackage[utf8]{inputenc}
\usepackage{amsmath}
\usepackage{amssymb}
\usepackage{graphicx}
\usepackage{graphics}
\usepackage{color}
\usepackage{dcolumn}
\usepackage{bm}
\usepackage[T1]{fontenc}
\usepackage{mathptmx}

\begin{document}


\title[S. Mumford et al.]{A Cantilever Torque Magnetometry Method for the Measurement of Hall Conductivity of Highly Resistive Samples}

\author{Samuel Mumford}
\affiliation{Geballe Laboratory for Advanced Materials, Stanford University, Stanford CA, 94305, USA}
\affiliation{Department of Physics, Stanford University, Stanford CA, 94305, USA}
\author{Tiffany Paul}%
\affiliation{Geballe Laboratory for Advanced Materials, Stanford University, Stanford CA, 94305, USA}
\affiliation{Department of Applied Physics, Stanford University, Stanford CA, 94305, USA}
\author{Seung Hwan Lee}
\affiliation{Department of Physics, Harvard University, Cambridge MA, 02138, USA}
\author{Amir Yacoby}
\affiliation{Department of Physics, Harvard University, Cambridge MA, 02138, USA}
\author{Aharon Kapitulnik}%
\affiliation{Geballe Laboratory for Advanced Materials, Stanford University, Stanford CA, 94305, USA}
\affiliation{Department of Physics, Stanford University, Stanford CA, 94305, USA}
\affiliation{Department of Applied Physics, Stanford University, Stanford CA, 94305, USA}%

\date{August 28, 2019}

\begin{abstract}
We present the first measurements of Hall conductivity utilizing a new torque magnetometry method designed for insulators. A Corbino disk exhibits a magnetic dipole moment proportional to Hall conductivity when voltage is applied across a test material. This magnetic dipole moment can be measured through torque magnetometry. The symmetry of this contactless technique allows for the measurement of Hall conductivity in previously inaccessible materials. Finally, a low-temperature noise bound, the lack of systematic errors on dummy devices, and a measurement of the Hall conductivity of sputtered indium tin oxide demonstrate the efficacy of the technique.
\end{abstract}
\maketitle

\section{\label{sec:intro}Introduction}

\begin{figure}
\centering
\includegraphics[width=1.0\columnwidth]{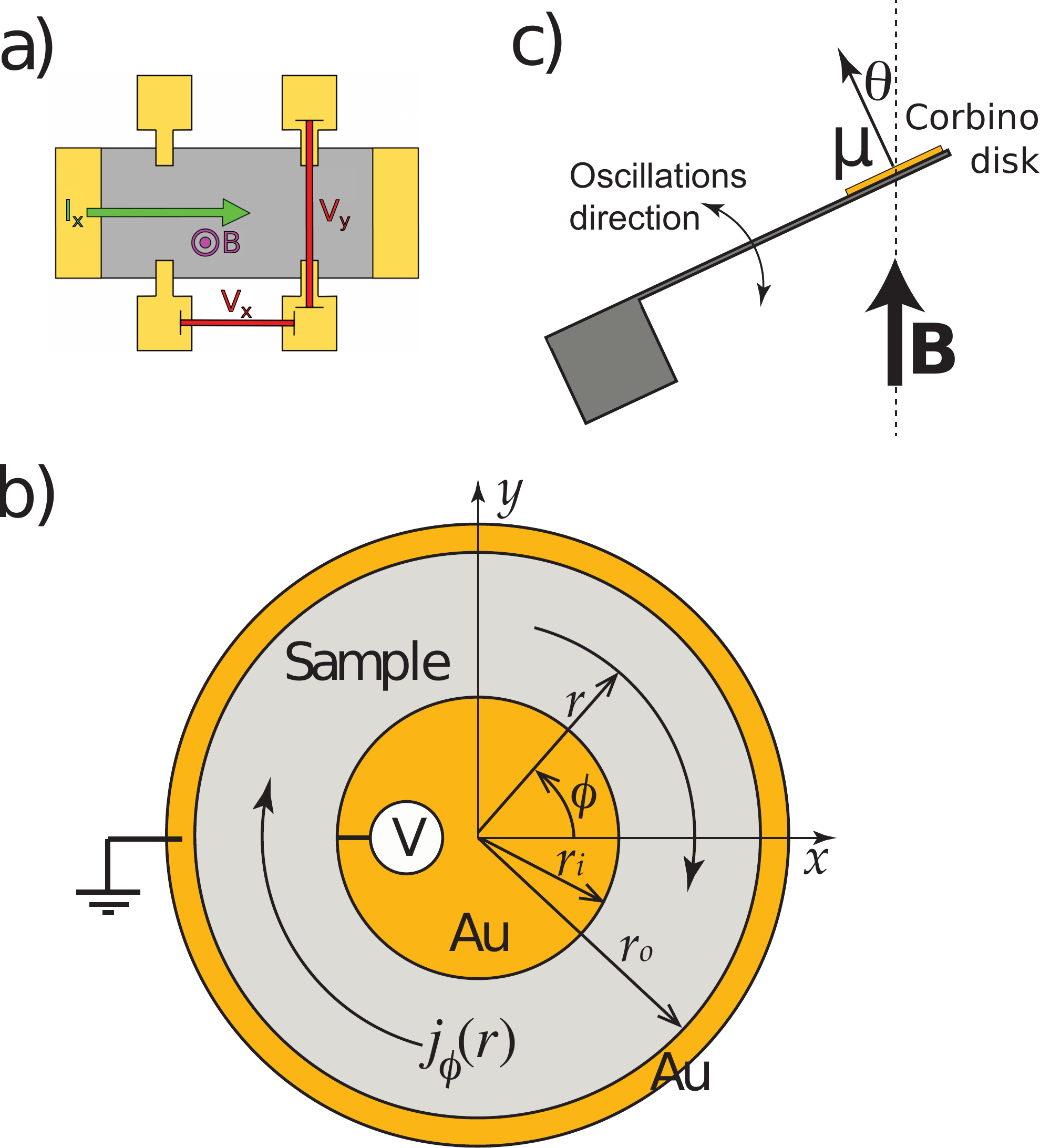}
\caption{a) The typical Hall bar consists of four contacts and a drive current $I$. A symmetry-breaking magnetic field $B$ allows for a non-diagonal terms in the resistivity tensor $\rho$. Correspondingly, there is a Hall voltage $V_{y} \propto \rho_{xy}I_{x}$ across contacts separated $\perp$ to $I_{x}$ as well as the longitudinal voltage $V_{x} \propto \rho_{xx}I_{x}$. \label{HBar} b) Corbino disk configuration used for $\sigma_{xy}$ measurements. Here the Au metallic contacts serve as the equipotential rings. c) Side view of torque magnetometry.}
 \label{ring}
\end{figure}

Measurements of transverse transport properties such as the Hall effect, Nernst effect, and transverse thermal conductivity have become of great importance in understanding modern quantum materials. However, such measurements are often made difficult, or even impossible, due to contamination of longitudinal transport effects. For example, in a standard Hall bar measurement of the Hall effect, the transverse voltage $V_{y}(H)$ is measured in response to the application of a longitudinal current $I_x$ in the presence of a perpendicular magnetic field $H\hat{z}$ using two contacts on opposite sides of the sample (see Fig.~\ref{ring}a). A common procedure to eliminate contributions from the longitudinal magnetoresistance due to contact misalignment invokes the odd symmetry of the effect to find Hall resistance $\rho_{xy}=[V_{y}(H)-V_{y}(-H)]/2I_x$. Here we use 2D notation where thickness is fixed. However, this simple procedure often fails when $\rho_{xx} \gg \rho_{xy}$, as is the case in the variable range hopping (VRH) regime of disordered insulators \cite{Mott1969, Shklovskii1984}, or on the insulating side of superconductor-insulator transition \cite{Steiner2005,Sambandamurthy2004, Paalanen1992, PhysRevLett.65.923, cao2018correlated}. Similarly, only a handful of Hall measurements were done in the VRH regime (see e.g. \cite{Hopkins1989,Koon1990}) despite detailed theories \cite{Pollak1981,Galperin1991}, and in general measurements were restricted to the vicinity of the metal-insulator transition. The issue is complicated further if one is interested in the transverse conductivity $\sigma_{xy}$, which can be calculated from the resistivity tensor, but with uncontrolled error-bars if $\rho_{xx}$ diverges. 
A direct measurement of  $\sigma_{xy}$ is needed to probe a variety of topological states of matter in the bulk of the material-system where edge states may dominate the transport. For example, standard transport approaches to measure the quantum Hall effect (QHE) in two dimensional electron gas (2DEG) interact directly with the edge states, with no ability to explore the existence of Hall currents in the bulk of the sample\cite{CMom}. Indeed, the original theoretical approach to explain the QHE by Laughlin \cite{Laughlin1981} used a closed metallic ribbon configuration, equivalent to a Corbino disk \cite{Corbino},  to demonstrate the effect. 

In this paper we demonstrate a new method for measuring $\sigma_{xy}$ in a Corbino disk configuration, where the induced Hall currents in the disk create a magnetic dipole moment that is measured by torque magnetometry. 
A circularly symmetric Corbino disk is shown in Fig.~\ref{ring}b. Fabricated at the end of a cantilever, it forms the basis of this $\sigma_{xy}$ measurement technique \cite{Corbino}. Applying a voltage $V$ between the inner and outer contacts creates a radial electric field $E_r$, which induces a circulating Hall current with a current density $j_{\phi}(r)$. This Hall current creates a  magnetic dipole moment $\mu$ parallel to the ring normal, which can be directly evaluated by
\begin{equation}
    \mu = \int j_{\phi}(r)\pi r^2 dr =\int \sigma_{xy} E_r\pi r^2 dr. \equiv \sigma_{xy}GV
\end{equation}
where $G$ is a geometrical factor. For concentric rings one obtains
\begin{equation}\label{mdm}
    \mu = \sigma_{xy}\frac{\pi  (r_{o}^2 - r_{i}^2)}{2 \ln(r_{o}/r_{i})}V
\end{equation}
where $r_i$ and $r_o$ are the inner and outer radii of the test material respectively. While real fabricated devices may deviate slightly from concentric rings, an image of the device can be used to numerically correct for that error. 

The magnetic dipole moment is then measured by means of torque magnetometry as shown in Fig.~\ref{ring}c, which allows for a high-precision contactless measurement. The dipole moment is measured without placing elements in series with the Hall current, and the torque measurement is insensitive to higher order magnetic moments caused by misalignment. The magnetic dipole moment of the full Corbino disk is also relatively insensitive to local disorder sources. Moreover, as the Corbino disk torque must be linear in $V$ and even in $B$, one may separate the signal due to the Hall effect from other effects due to cantilever heating or longitudinal current by signal symmetry. 


\section{\label{sec:meth}Methods}

\subsection{\label{sec:meas}Measurement Concept}

Cantilever torque magnetometry utilizes a high-$Q$ resonator to detect the interaction between a magnetic dipole and an external magnetic field~\cite{PERFETTI2017171, PhysRevB.64.014516, Bleszynski-Jayich272}. The angular response $\theta$ of a cantilever with moment of inertia $A$, resonant frequency $\omega_{0}=2\pi f_0$, and quality factor $Q$ subject to an external torque $\tau$ may be approximated as a damped harmonic oscillator following~\cite{s8053497}
\begin{equation}\label{baseEQ}
    A\ddot{\theta} + QA \omega_{0}\dot{\theta} + A\omega_{0}^2 \theta = \tau.
\end{equation}
An external magnetic field $\vec{B}$ exerts a torque~\cite{PERFETTI2017171}
\begin{equation}
    \vec{\tau} = \vec{\mu} \times \vec{B}.
\end{equation}
If the dipole moment and magnetic field are aligned in the cantilever equilibrium position, an effective detuning torque
\begin{equation}\label{detune}
    \tau_{D} = \mu B\sin(\theta) \approx \mu B\theta
\end{equation}
results as the cantilever oscillates. Inserting $\tau_{D}$ into Equation~\ref{baseEQ} shifts the resonant frequency by
\begin{equation}\label{fu}
    A\omega_{0}^2 \rightarrow A\omega_{0}^2 - \mu B\textrm{,\ \ \ \ \ \  or \ \ \ \ \ \  } \frac{\Delta \omega_{0}}{\omega_{0}} = \frac{\mu B}{2A\omega_{0}^2}.
\end{equation}
Using Eqn.~\ref{mdm}, the shift in resonant frequency can be related to the applied voltage, magnetic field, and $\sigma_{xy}$ by
\begin{equation}
\label{hallShift}
\delta f_0=\frac{GV}{8\pi^2Af_0}B\sigma_{xy}.
\end{equation}
Measurement of changes in $f_{0}$ of a patterned cantilever with voltage therefore probes $\sigma_{xy}$ without polluting terms from $\rho_{xx}$.

\subsection{\label{sec:fab}Device Fabrication}

\begin{figure}
\centering
\includegraphics[width=1.0\columnwidth]{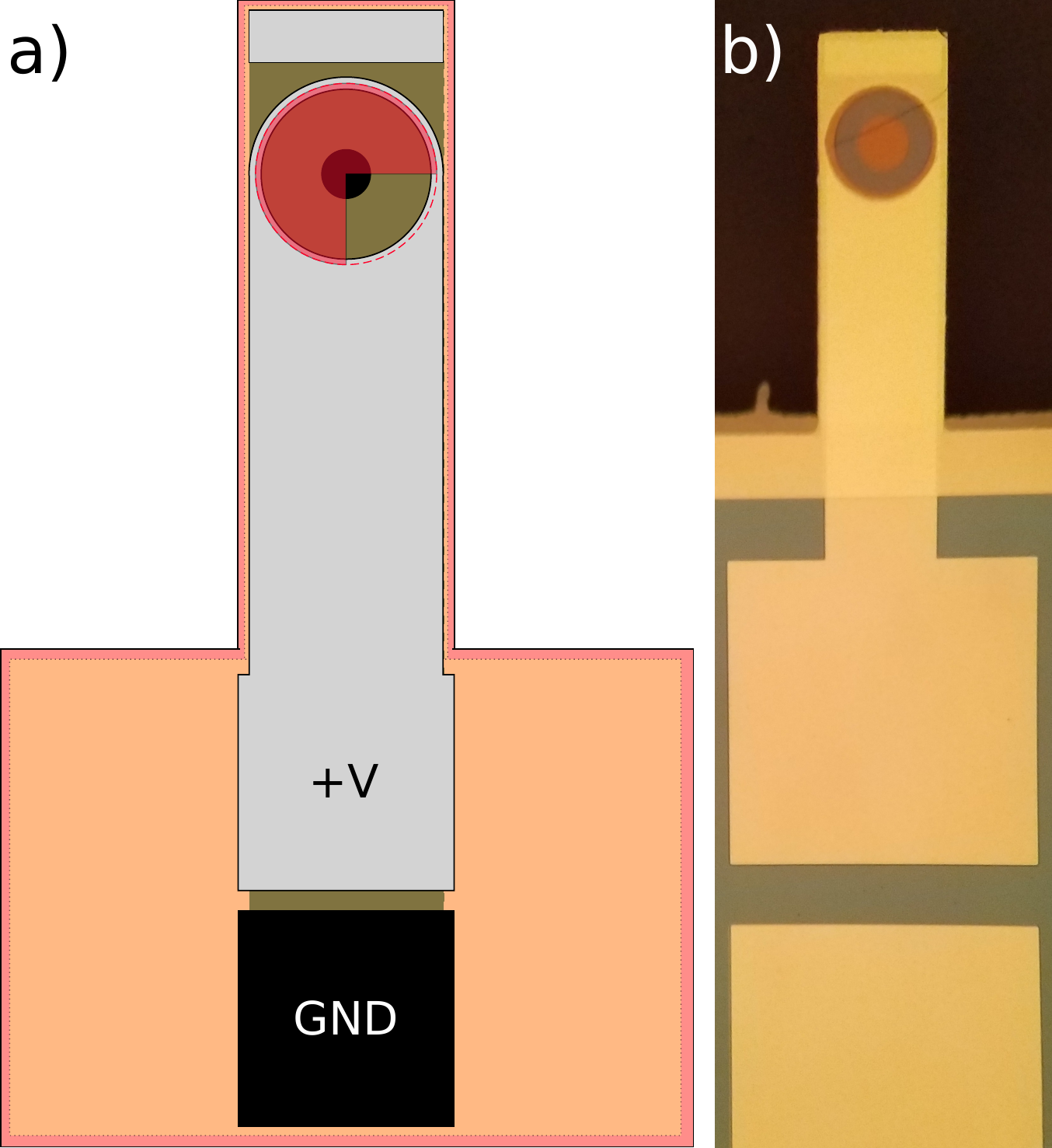}
\caption{ \label{Cartoon} a) Schematic drawing of a cantilever patterned with a Corbino disk in the planar coaxial design. An insulating barrier (orange) separates the Pt inner contact (black) and the Pt outer ring contact (gray). The test material (red) is deposited in a circle connecting the voltage contacts and the underlying Si is shown in pink. b) A planar coaxial Corbino disk cantilever with ITO as the test material.}
\end{figure}

Corbino disks were patterned on high-$Q$ single-crystal silicon cantilevers as shown in Fig.~\ref{Cartoon}. Fabrication was performed using photolithography as all features are larger than 15~$\mu$m. Fabrication began with a silicon on insulator (SOI) wafer with a 450~$\mu$m handle layer, a 4~$\mu$m buried oxide layer, and a 2 or 3~$\mu$m device layer of (001) Si depending on the design. A CVD SiO$_{2}$ layer was first deposited onto the handle side and this oxide was plasma etched in an array of square windows. Next, 25~nm of Ti-Pt was patterned in the shape of the cantilever on the device side. This conductive layer serves as a ground plane and separates the Corbino disk voltages from the underlying Si. A 40~nm thick barrier of ALD HfO$_{2}$ and CVD nitride were then grown on top of the Pt to electrically separate the grounding plane from subsequently deposited layers. On the planar coaxial design shown in Fig.~\ref{Cartoon}, a hole was etched in this insulating layer for the grounded central Corbino disk contact and the Pt outer contact was deposited directly onto the SiN-HfO$_{2}$ insulating layer. The ground plane serves both as the inner contact and as a conductive barrier in the planar coaxial design. The cantilever shape was etched out from the device layer Si using a Bosch etcher. Finally, the test material was deposited between the inner and outer contacts and the cantilever was released using a backside Bosch etch and a final oxide plasma etch.

For devices with separate grounded wires on top of the ground plane, such as the Ge dummy device of Section \ref{sec:Ge}, the SiN-HfO$_2$ insulating barrier was not etched. A thin Ti-Au inner contact was instead deposited onto this insulating substrate and a 20~nm thick ALD HfO$_{2}$ ring was patterned to cover most of this inner contact. This ring separates the inner and outer voltage contacts. The resulting cantilevers with separate ground planes are 250~$\mu$m x 600~$\mu$m x 3~$\mu$m and $f_{0} \sim 10$~kHz. The planar coaxial design cantilevers are  200~$\mu$m x 600~$\mu$m x 2~$\mu$m and with $f_{0} \sim 7$~kHz. Both exhibit $Q \sim 25000$ at pressures below $1\times10^{-4}$~Torr.

\subsection{\label{sec:detect}Interferometric Resonant Frequency Detection}

\begin{figure}
\centering
\includegraphics[width=1.0\columnwidth]{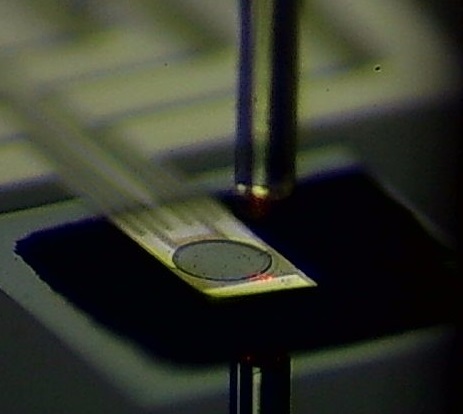}
\caption{ \label{Inter} A cleaved fiber above a cantilever forming an interferometer. The two interfering light sources are reflected light from the fiber end and the cantilever surface. The fiber was aligned with a three-axis stage and a red laser before being epoxied to the cantilever wafer. A separate pad of Pt was patterned on the end of the cantilever for alignment.}
\end{figure}

The resonant frequency of the cantilever is tracked with a fiber interferometer. The output of a 1310~nm fiber-coupled laser diode is first fed through a 90-10 splitter. The majority of the laser power goes to a reference photodiode and the remaining 10$\%$ of laser power is connected to a cleaved fiber optic cable. The cleaved fiber end is then aligned over a cantilever to form an interferometer as shown in Fig.~\ref{Inter}. The output from this interferometer is converted into a voltage by a photodiode and computer processed after analog-to-digital conversion. The interferometer voltage $V(t)$ for laser wavelength $\lambda$, peak to peak voltage $V_{pp}$, and fiber-cantilever distance $\Delta z$ is
\begin{equation}\label{inter}
    V(t) \approx \frac{2\pi V_{pp} \Delta z(t)}{\lambda}\sin(\frac{4\pi \Delta z_{0}}{\lambda}) \propto \Delta z(t).
\end{equation}
The fiber interferometer thus provides the means to precisely track cantilever motion.

\begin{figure}
\centering
\includegraphics[width=1.0\columnwidth]{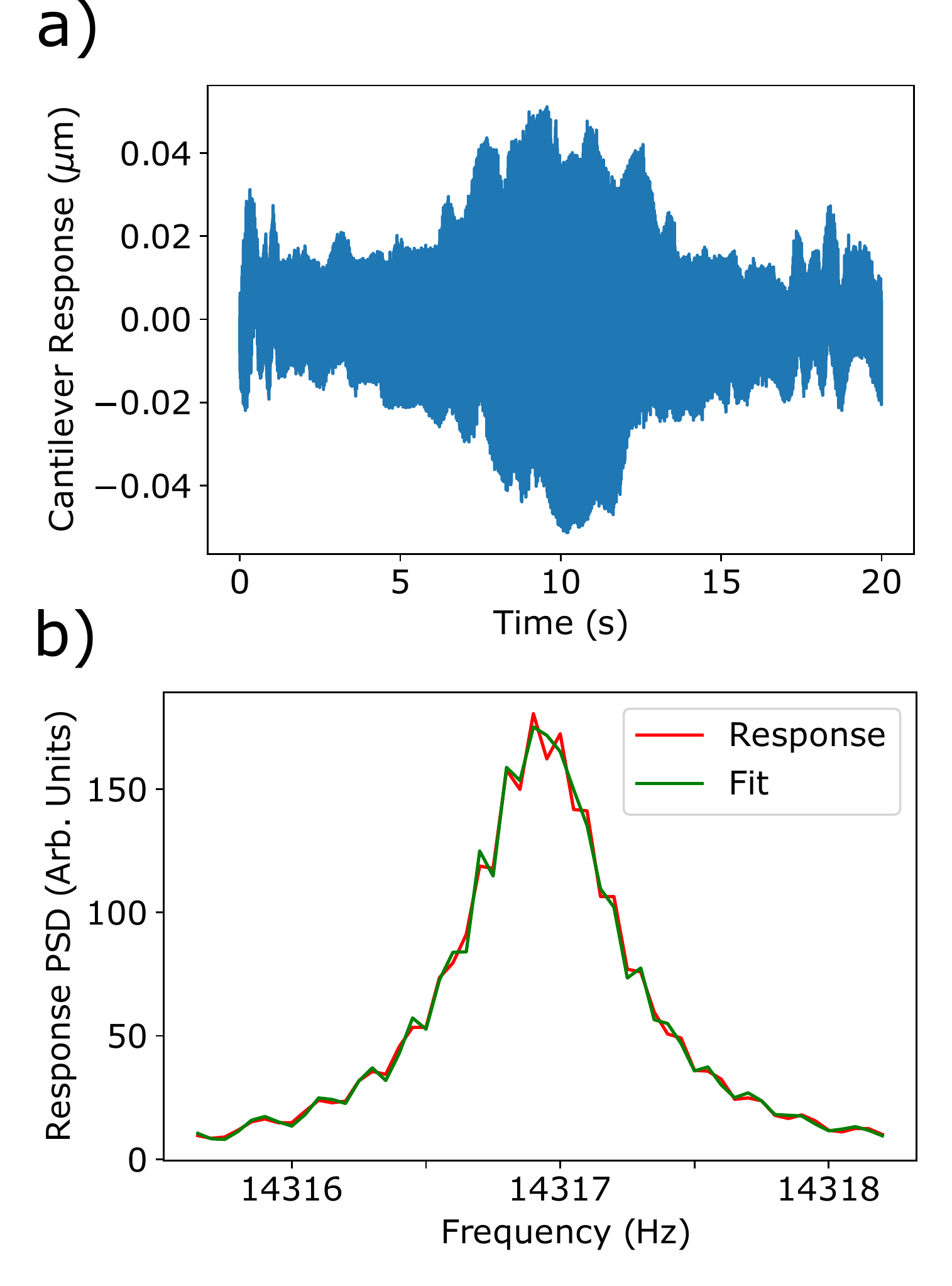}
\caption{a) Real time response of a 3$~\mu$m thick Si cantilever to a sweep drive. The amplitude of driving power modulation was 50~$\mu$W. Each drive and fit was performed over 20~s so subsequent drives would be independent with a cantilever response time of $\sim 2$~s. b)~Power spectral density of the cantilever response seen in part a. Response is fit to a damped harmonic oscillator equation finding $f_{0}$ of 14316.9~Hz and $Q$ of 23000.}
\label{response}
\end{figure}

Cantilever resonant frequency is measured by observing the response to a radiation pressure drive. A 1550~nm fiber coupled laser is connected to the cleaved fiber used for interferometry. The power of this laser is reflected off of a Pt pad near the end of the cantilever. This laser power is modulated with a frequency sweeping voltage over a frequency width $\delta f$ and time $t_{sw}$
\begin{equation}
    V_{sw} = V_{0}\sin[2\pi(f_{0} - \frac{\delta f}{2} + \frac{\delta f t}{2 t_{sw}})t]
\end{equation}
to create a sweeping radiation pressure drive. The sweeping drive voltage is also connected to a reference port in the analog-to-digital converter to fit for the cantilever response function, shown in Fig.~\ref{response}.

As shown in Appendix \ref{sec:Noise}, the minimum detectable shift in resonant frequency for a cantilever with length $L$  and vibrational temperature $T$ driven for time $t_{samp}$ is
\begin{equation}\label{Fun}
    \frac{\Delta \omega_{0}}{\omega_{0}} = \frac{2L}{\lambda}\sqrt{\frac{2k_{b}T\pi}{A\omega_{0}^3t_{samp} Q}},
\end{equation}
which by Equations \ref{mdm} and \ref{fu} translates to a minimum detectable $\sigma_{xy}$ of 
\begin{equation}\label{noisespec}
    \delta \sigma_{xy} = \frac{4L \ln(r_{o}/r_{i})}{\lambda(r_{o}^2 - r_{i}^2)V B}\sqrt{\frac{2k_{b}TA\omega_{0}}{\pi t_{samp} Q}}.
\end{equation}
For this cantilever design, dilution refrigerator temperature of 0.1~K, a 1~T magnet, and 0.1~V applied, the minimum detectable $\sigma_{xy} \sim 10^{-9}~ \Omega^{-1}$. Such uncertainty improves upon Hall bar measurements for insulating samples extrapolated to $T \rightarrow 0$ by a factor of >~$10^{5}$~\cite{Breznay2016}. Note as sample heating scales as $V^2/\rho$, the minimum observable $\sigma_{xy}$ scales as $1/\sqrt{\rho}$ and decreases dramatically for insulators at low temperature.

\section{\label{sec:exp}Experimental Results}

\subsection{\label{sec:dummy}Systematics Tests: Conductive Pt Device}

\begin{figure}
\centering
\includegraphics[width=1.0\columnwidth]{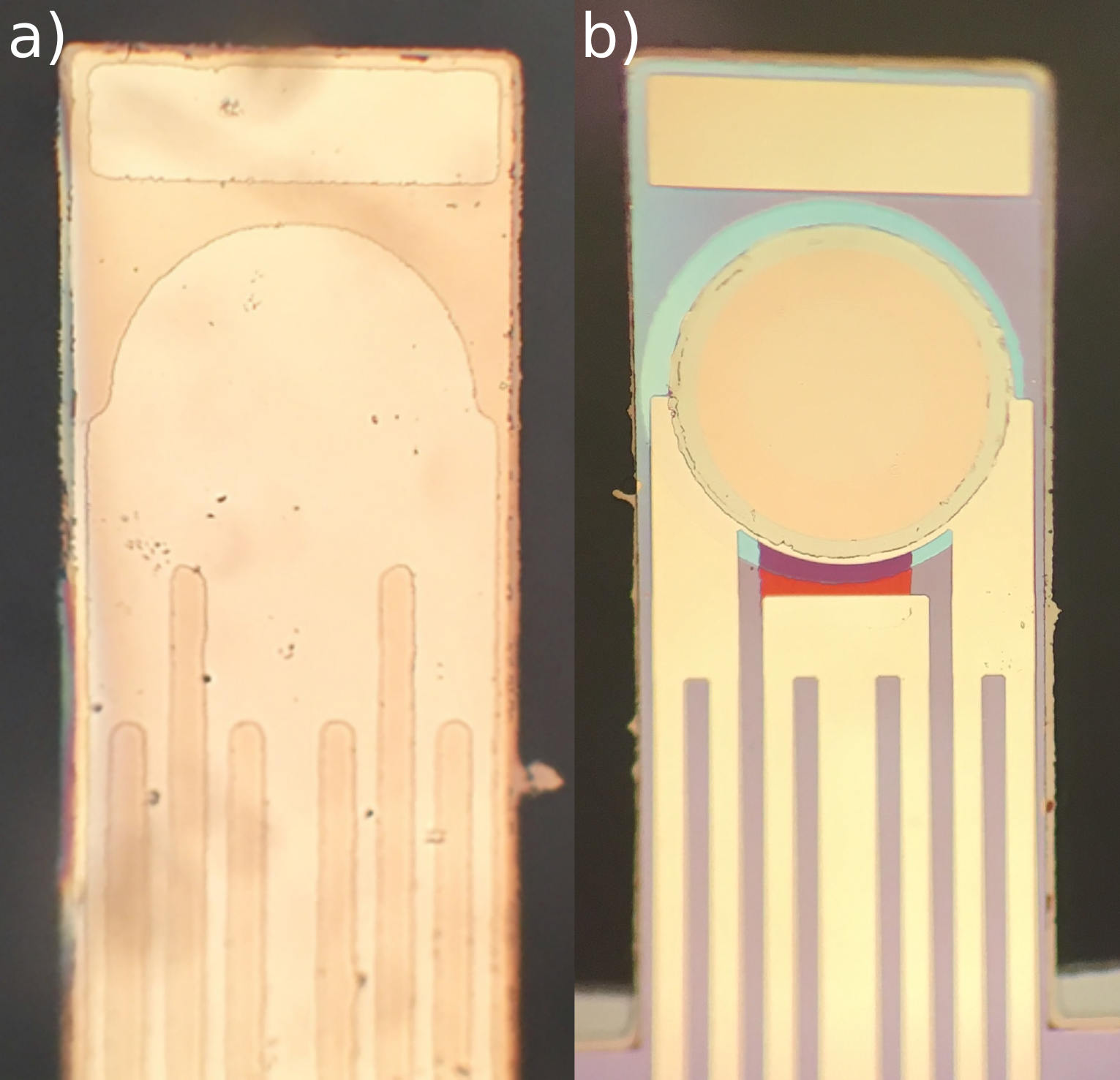}
\caption{ \label{ZO} a) A dummy device with Pt wires used to look for systematic errors and noise bounds. b) A Corbino disk cantilever with Ge as a test material. Note the pairs of wires for the inner and outer contacts to allow for four wire measurement of the voltage across the disk.}
\end{figure}

Dummy devices without Corbino disks were investigated to check for systematic errors which could create a shift in $f_{0}$ similar to a Hall signal. An initial dummy device consisted of Pt wires and a ground plane patterned on a 3$~\mu$m thick single-crystal Si cantilever, shown in Fig.~\ref{ZO}a. The resistance of the Pt wires was 850~$\Omega$, and alternating voltages $V = \pm~350$~mV were applied across the device to look for an odd in $V$ and even in $B$ shift in $f_{0}$. The observed shift in $f_{0}$ at room temperature between $\pm~ V$, $f_{0}(+V) - f_{0}(-V)$ or $\delta f_{0}$, is $-4\pm26~\mu$Hz. A second Pt dummy cantilever was tested at room temperature and similarly displays $\delta f_{0} = 29 \pm 28 ~\mu$Hz. The grounded Pt cantilevers thus show no evidence of a zero-field $\delta f_{0}$ as expected. The Pt dummy was also cooled to 4.2~K and exposed to $\pm~1$~T of magnetic field in a Janis SVT cryostat. The sum of $\delta f_{0}$ for $\pm~B$ is $0.27 \pm 0.21~$mHz and the zero field $\delta f_{0}$ at 4.2~K is -$0.03 \pm 0.1~$mHz. The Pt dummy cantilevers thus demonstrate no evidence of a zero-field or Hall-like $\delta f_{0}$ at both room temperature and liquid helium temperature.

\subsection{\label{sec:Ge}Systematics Tests: Insulating Ge Device}

Full Corbino disk cantilevers with evaporated amorphous Ge as a test material were also fabricated (Fig.~\ref{ZO}b). The insulating Ge serves both as a test for spurious $\delta f_{0}$ with voltage applied across the ring and as a means to verify that the full fabrication procedure did not create unintended electrical connections. The 3~$\mu$m thick Ge devices are both electrically and mechanically viable, with resistances of $>20~$M$\Omega$ and $Q\sim 25000$. Once more, there is no evidence of a Hall-like shift in $f_{0}$ at room temperature, finding $\delta f_{0} = 88 \pm 91~\mu$Hz in 0 field and $\delta f_{0} = 30 \pm 80~\mu$Hz above a 0.3~T static magnet.

In conclusion, the dummy device tests demonstrate that design flaws creating Hall-like $\delta f_{0}$ have been eliminated. Hall signal therefore may be distinguished by voltage and magnetic field behavior from other shifts in $f_{0}$ in patterned Si cantilevers with Corbino disks.

\subsection{Measurements of ITO Corbino Disk}

Corbino cantilevers with indium tin oxide (ITO) as a sample material were fabricated for first measurements of Hall signal. ITO was chosen as an example of a disordered itinerant system which can be tuned through a metal-insulator transition (MIT) by changing the tin and oxygen content. Here we sputtered 50~nm of ITO with resistivity $3.5~\times~10^{-3}~\Omega$-cm. Such ITO should exhibit $\sigma_{xy}~>~1 \times 10^{-7}~\Omega^{-1}$ at 5~T~\cite{kurdesau_khripunov_cunha_kaelin_tiwari_2006}. Based on the observed carrier density in this system such ITO should be in the vicinity of the MIT with $k_F\ell \lesssim 1$. With typical carrier density for this material, it is expected that $\rho_{xy}/\rho_{xx}\sim10^{-4}$ which is on the borderline of standard methods of Hall effect detection\cite{kurdesau_khripunov_cunha_kaelin_tiwari_2006}.

\begin{figure}
\centering
\includegraphics[width=1.0\columnwidth]{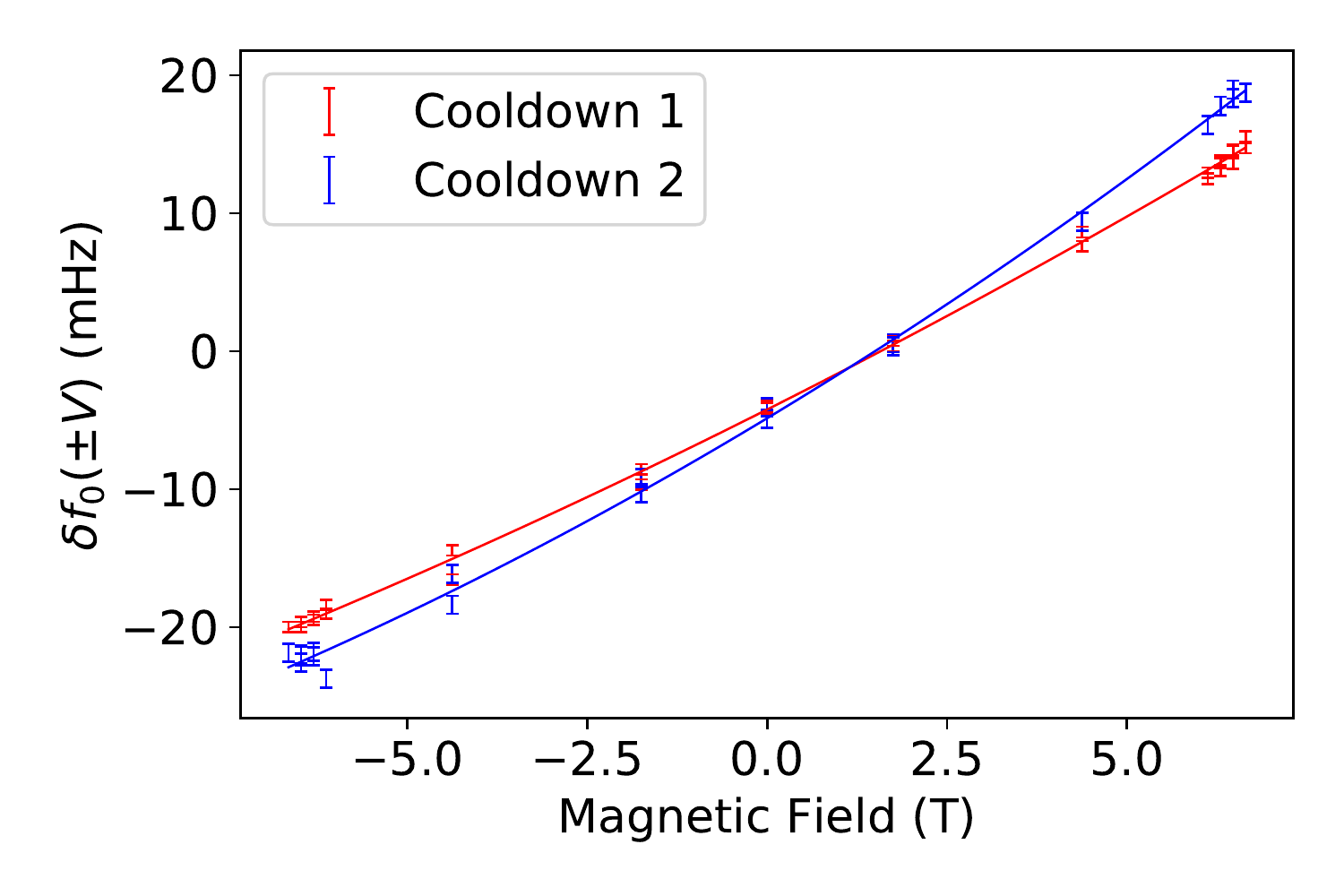}
\caption{ \label{Odd} Data and fit for $\delta f_{0}(B)$ with $\pm~0.1~V$ applied across the first ITO cantilever in two cooldowns. Note the overriding linear dependence due to patterning asymmetry of the current-carrying Pt wires. This can be used to compare $A$ between cooldowns on the same cantilever.}
\end{figure}

\begin{figure}
\centering
\includegraphics[width=1.0\columnwidth]{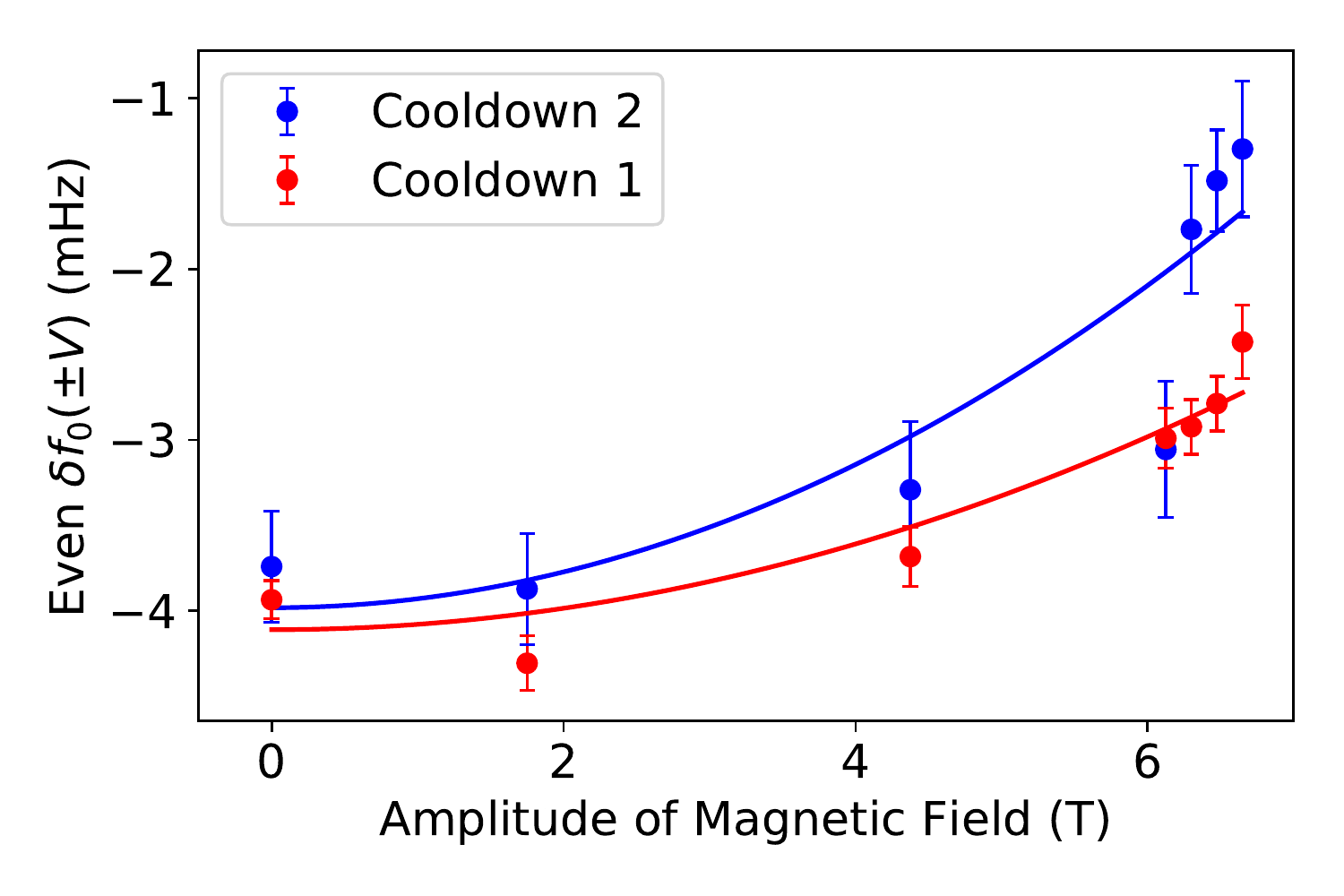}
\caption{ \label{Resid} Data and quadratic fit for the even component of $\delta f_{0}(B)$. There is a clear quadratic Hall signal in all cooldowns with sputtered ITO as a test material.}
\end{figure}

Cantilevers with ITO Corbino disks were tested to verify a real Hall signal can be seen. In an effort to improve future torque sensitivity, thinner 2$~\mu$m thick planar coaxial cantilevers were fabricated as shown in Fig.~\ref{Cartoon}. Fig.~\ref{Odd} shows $\delta f_0 (B)$ of a cantilever in two datataking runs at $4.2~$K. After fitting to a second order polynomial and subtracting the linear component, the Hall signal of the ITO can clearly be observed as a quadratic dependence in $\delta f_{0}(B)$ in Fig.~\ref{Resid}. This quadratic dependence is seen in both datataking runs, yielding fit coefficients of $29.2 \pm 4.6 ~\mu$Hz/T$^{2}$ and $47.1 \pm 9.3 ~\mu$Hz/T$^{2}$. The order of datataking was reversed for the second run, confirming that the quadratic dependence in $\delta f_{0}(B)$ is not a temperature or time effect. A second cantilever from the same SOI wafer was tested to confirm that the curvature of $\delta f_{0}(B)$ could be attributed to the ITO. On the second device the curvature in $\delta f_{0}(B)$ is $46.7 \pm 8.7~\mu$Hz/T$^{2}$. Using Eqn. \ref{hallShift}, at 5~T the quadratic fit coefficient of each cantilever translates to $\sigma_{xy}$ of $(2.33 \pm 0.40) \times10^{-7}~\Omega^{-1}$ and $(2.21\pm 0.42) \times10^{-7}~\Omega^{-1}$. Converting back to resistivities, $\rho_{xy}~\sim~0.1~\Omega$ or $5\times10^{-7}~\Omega$-cm in 3D units. This statistically consistent result for $\sigma_{xy}$ across different cantilevers, cooldowns, and data taking procedures is in agreement with previous measurements of sputtered ITO and verifies that the observed quadratic dependence in $\delta f_{0}(B)$ is caused by the ITO Corbino disk. The small ratio of $\rho_{xy}/\rho_{xx}$ also demonstrates the effectiveness of the technique.

\section{Summary}

In conclusion, Corbino disk torque magnetometry is a viable method for measuring $\sigma_{xy}$. First, the initial challenge of fabricating high-$Q$ cantilevers with patterned Corbino disks and contacts has been completed. The resonant frequency of such devices can be measured with a fractional uncertainty of less than one part in 10$^7$, even in conditions that have not yet employed vibration isolation, high vacuum conditions, or special shielding. The Corbino disk cantilevers have also been tested for errors in fabrication, data collection procedures, and analysis protocols. This allowed for the elimination of systematic errors and spurious signals, both when applying current and voltage across the disk. These new cantilevers have also been used to measure $\sigma_{xy}$ of sputtered ITO with nominal resistivity of $\rho_{xx}~\sim~3.5~\times~10^{-3}~\Omega$-cm, demonstrating the ability to detect the Hall effect in samples where $\rho_{xy}/\rho_{xx}\sim~10^{-4}$ \cite{kurdesau_khripunov_cunha_kaelin_tiwari_2006}, which is generally difficult to measure using standard techniques.


\begin{acknowledgments}
This work was funded by the Army Research Office grant W911NF1710588, and by the Gordon and Betty Moore Foundation through Emergent Phenomena in Quantum Systems (EPiQS) Initiative Grant GBMF4529. This work was also funded in part by a QuantEmX grant from ICAM and the Gordon and Betty Moore Foundation through Grant GBMF5305 to Seung Hwan Lee.
\end{acknowledgments}
\bigskip
\appendix

\section{\label{sec:Noise}Theoretical Noise Floor Calculation}
    For a cantilever response function $R(\omega, \omega_{0})$ to an applied torque $\tau_{app}$, the observed change in cantilever oscillation when the resonant frequency shifts by $\Delta \omega_{0}$ is
    \begin{equation*}
        \Delta\theta_{sig}(\omega) = \tau_{app}(\omega)\frac{dR}{d\omega_{0}}\Delta\omega_{0}.
    \end{equation*}
    As seen in Fig.~\ref{response}a, $\tau_{app}$ is limited by interferometer wavelength $\lambda$. For a cantilever of length $L$ the maximum angle of deflection over time $t_{samp}$ is
    \begin{equation*}
        \Delta \theta_{max} = \lambda/2L = \tau_{max}(\omega)R(\omega)2\pi/t_{samp}.
    \end{equation*}
    The largest possible signal for a single-frequency drive therefore is
    \begin{equation*}
        \Delta\theta_{sig}(\omega) = \frac{\lambda t_{samp}}{2L}\frac{dR}{d\omega_{0}}\frac{1}{R(\omega)}\Delta\omega_{0}.
    \end{equation*}
    
    The fundamental experimental noise source is thermal vibration. By equipartition $k_BT = A\langle\theta^2(t)\rangle$, or assuming a white noise thermal drive $\tau_{therm}$
    \begin{equation*}
        \langle\theta^2(t)\rangle = \frac{\tau^2_{therm}}{t_{samp}}\int R^2(\omega) d\omega = \frac{k_BT}{A}.
    \end{equation*}
    So with
    \begin{equation*}
        \tau_{therm}(\omega) = \sqrt{2k_BTA\omega_{0}\frac{t_{samp}}{\pi Q}},
    \end{equation*}
    the noise response is
    \begin{equation*}
        \Delta\theta(\omega) = \tau_{therm}(\omega)R(\omega, \omega_{0}, A, Q).
    \end{equation*}
    Setting the signal to noise ratio to 1, the minimum detectible frequency shift is
    \begin{equation*}
        \Delta\omega_{0} = \frac{4L}{\lambda t_{samp}}\sqrt{2\pi k_BTA\omega_{0}\frac{t_{samp}}{ Q}}\min(|(\frac{dR}{d\omega_{0}})^{-1}R^2(\omega)|),
    \end{equation*}
    or 
    \begin{equation*}
        \frac{\Delta \omega_{0}}{\omega_{0}} = \frac{2L}{\lambda}\sqrt{\frac{2k_BT\pi}{A\omega_{0}^3t_{samp} Q}}.
    \end{equation*}
    
    This unitless noise bound has a simple physical explanation. Using that $\lambda/2L$ is the maximum angle of the driven cantilever, that $\omega_{0}t_{samp}/2\pi$ is averaging time counted in number of oscillations, and that narrower resonances will have less uncertain $\omega_{0}$, the noise bound is truly
    \begin{equation*}
        \frac{\Delta \omega_{0}}{\omega_{0}} = \sqrt{\frac{\textrm{Thermal Energy}}{\textrm{Driven Energy}*\textrm{Time in Oscillations}*Q}}.
    \end{equation*}
    Finally, using Equations \ref{fu} and \ref{mdm},
    \begin{equation*}
        \delta \sigma_{xy} = \frac{4L \ln(r_{o}/r_{i})}{\lambda(r_{o}^2 - r_{i}^2)V B}\sqrt{\frac{2k_BTA\omega_{0}}{\pi t_{samp} Q}}.
    \end{equation*}
    
    The theoretical uncertainty bound can be compared to the present dummy cantilever data. Calculating $A$ by observing the response magnitude to a known drive and using the vibrational temperature from equipartition along with Equation~\ref{Fun}, the best possible fractional uncertainty for such an experiment should be $\sim 10^{-11}$. The fractional uncertainty without vibration isolation is currently $\sim 10^{-9}$.

\nocite{*}
\bibliography{bibl.bib}

\begin{thebibliography}{20}%
\makeatletter
\providecommand \@ifxundefined [1]{%
 \@ifx{#1\undefined}
}%
\providecommand \@ifnum [1]{%
 \ifnum #1\expandafter \@firstoftwo
 \else \expandafter \@secondoftwo
 \fi
}%
\providecommand \@ifx [1]{%
 \ifx #1\expandafter \@firstoftwo
 \else \expandafter \@secondoftwo
 \fi
}%
\providecommand \natexlab [1]{#1}%
\providecommand \enquote  [1]{``#1''}%
\providecommand \bibnamefont  [1]{#1}%
\providecommand \bibfnamefont [1]{#1}%
\providecommand \citenamefont [1]{#1}%
\providecommand \href@noop [0]{\@secondoftwo}%
\providecommand \href [0]{\begingroup \@sanitize@url \@href}%
\providecommand \@href[1]{\@@startlink{#1}\@@href}%
\providecommand \@@href[1]{\endgroup#1\@@endlink}%
\providecommand \@sanitize@url [0]{\catcode `\\12\catcode `\$12\catcode
  `\&12\catcode `\#12\catcode `\^12\catcode `\_12\catcode `\%12\relax}%
\providecommand \@@startlink[1]{}%
\providecommand \@@endlink[0]{}%
\providecommand \url  [0]{\begingroup\@sanitize@url \@url }%
\providecommand \@url [1]{\endgroup\@href {#1}{\urlprefix }}%
\providecommand \urlprefix  [0]{URL }%
\providecommand \Eprint [0]{\href }%
\providecommand \doibase [0]{http://dx.doi.org/}%
\providecommand \selectlanguage [0]{\@gobble}%
\providecommand \bibinfo  [0]{\@secondoftwo}%
\providecommand \bibfield  [0]{\@secondoftwo}%
\providecommand \translation [1]{[#1]}%
\providecommand \BibitemOpen [0]{}%
\providecommand \bibitemStop [0]{}%
\providecommand \bibitemNoStop [0]{.\EOS\space}%
\providecommand \EOS [0]{\spacefactor3000\relax}%
\providecommand \BibitemShut  [1]{\csname bibitem#1\endcsname}%
\let\auto@bib@innerbib\@empty
\bibitem [{\citenamefont {Mott}(1969)}]{Mott1969}%
  \BibitemOpen
  \bibfield  {author} {\bibinfo {author} {\bibfnamefont {N.~F.}\ \bibnamefont
  {Mott}},\ }\href {\doibase 10.1080/14786436908216338} {\bibfield  {journal}
  {\bibinfo  {journal} {The Philosophical Magazine: A Journal of Theoretical
  Experimental and Applied Physics}\ }\textbf {\bibinfo {volume} {19}},\
  \bibinfo {pages} {835} (\bibinfo {year} {1969})}\BibitemShut {NoStop}%
\bibitem [{\citenamefont {Shklovskii}\ and\ \citenamefont
  {Efros}(1976)}]{Shklovskii1984}%
  \BibitemOpen
  \bibfield  {author} {\bibinfo {author} {\bibfnamefont {B.~I.}\ \bibnamefont
  {Shklovskii}}\ and\ \bibinfo {author} {\bibfnamefont {A.~L.}\ \bibnamefont
  {Efros}},\ }\href@noop {} {\emph {\bibinfo {title} {{Electronic Properties of
  Doped Semiconductors}}}},\ \bibinfo {edition} {1st}\ ed.\ (\bibinfo
  {publisher} {Springer-Verlag, Berlin Heidelberg},\ \bibinfo {year}
  {1976})\BibitemShut {NoStop}%
\bibitem [{\citenamefont {Steiner}\ and\ \citenamefont
  {Kapitulnik}(2005)}]{Steiner2005}%
  \BibitemOpen
  \bibfield  {author} {\bibinfo {author} {\bibfnamefont {M.}~\bibnamefont
  {Steiner}}\ and\ \bibinfo {author} {\bibfnamefont {A.}~\bibnamefont
  {Kapitulnik}},\ }\href {\doibase https://doi.org/10.1016/j.physc.2005.02.014}
  {\bibfield  {journal} {\bibinfo  {journal} {Physica C: Superconductivity}\
  }\textbf {\bibinfo {volume} {422}},\ \bibinfo {pages} {16 } (\bibinfo {year}
  {2005})}\BibitemShut {NoStop}%
\bibitem [{\citenamefont {Sambandamurthy}\ \emph {et~al.}(2004)\citenamefont
  {Sambandamurthy}, \citenamefont {Engel}, \citenamefont {Johansson},\ and\
  \citenamefont {Shahar}}]{Sambandamurthy2004}%
  \BibitemOpen
  \bibfield  {author} {\bibinfo {author} {\bibfnamefont {G.}~\bibnamefont
  {Sambandamurthy}}, \bibinfo {author} {\bibfnamefont {L.}~\bibnamefont
  {Engel}}, \bibinfo {author} {\bibfnamefont {A.}~\bibnamefont {Johansson}}, \
  and\ \bibinfo {author} {\bibfnamefont {D.}~\bibnamefont {Shahar}},\ }\href
  {\doibase 10.1103/PhysRevLett.92.107005} {\bibfield  {journal} {\bibinfo
  {journal} {Physical review letters}\ }\textbf {\bibinfo {volume} {92}},\
  \bibinfo {pages} {107005} (\bibinfo {year} {2004})}\BibitemShut {NoStop}%
\bibitem [{\citenamefont {Paalanen}\ \emph {et~al.}(1992)\citenamefont
  {Paalanen}, \citenamefont {Hebard},\ and\ \citenamefont
  {Ruel}}]{Paalanen1992}%
  \BibitemOpen
  \bibfield  {author} {\bibinfo {author} {\bibfnamefont {M.~A.}\ \bibnamefont
  {Paalanen}}, \bibinfo {author} {\bibfnamefont {A.~F.}\ \bibnamefont
  {Hebard}}, \ and\ \bibinfo {author} {\bibfnamefont {R.~R.}\ \bibnamefont
  {Ruel}},\ }\href@noop {} {\bibfield  {journal} {\bibinfo  {journal} {Phys.
  Rev. Lett.}\ }\textbf {\bibinfo {volume} {69}},\ \bibinfo {pages} {1604}
  (\bibinfo {year} {1992})}\BibitemShut {NoStop}%
\bibitem [{\citenamefont {Fisher}(1990)}]{PhysRevLett.65.923}%
  \BibitemOpen
  \bibfield  {author} {\bibinfo {author} {\bibfnamefont {M.~P.~A.}\
  \bibnamefont {Fisher}},\ }\href {\doibase 10.1103/PhysRevLett.65.923}
  {\bibfield  {journal} {\bibinfo  {journal} {Phys. Rev. Lett.}\ }\textbf
  {\bibinfo {volume} {65}},\ \bibinfo {pages} {923} (\bibinfo {year}
  {1990})}\BibitemShut {NoStop}%
\bibitem [{\citenamefont {Cao}\ \emph {et~al.}(2018)\citenamefont {Cao},
  \citenamefont {Fatemi}, \citenamefont {Demir}, \citenamefont {Fang},
  \citenamefont {Tomarken}, \citenamefont {Luo}, \citenamefont
  {Sanchez-Yamagishi}, \citenamefont {Watanabe}, \citenamefont {Taniguchi},
  \citenamefont {Kaxiras} \emph {et~al.}}]{cao2018correlated}%
  \BibitemOpen
  \bibfield  {author} {\bibinfo {author} {\bibfnamefont {Y.}~\bibnamefont
  {Cao}}, \bibinfo {author} {\bibfnamefont {V.}~\bibnamefont {Fatemi}},
  \bibinfo {author} {\bibfnamefont {A.}~\bibnamefont {Demir}}, \bibinfo
  {author} {\bibfnamefont {S.}~\bibnamefont {Fang}}, \bibinfo {author}
  {\bibfnamefont {S.~L.}\ \bibnamefont {Tomarken}}, \bibinfo {author}
  {\bibfnamefont {J.~Y.}\ \bibnamefont {Luo}}, \bibinfo {author} {\bibfnamefont
  {J.~D.}\ \bibnamefont {Sanchez-Yamagishi}}, \bibinfo {author} {\bibfnamefont
  {K.}~\bibnamefont {Watanabe}}, \bibinfo {author} {\bibfnamefont
  {T.}~\bibnamefont {Taniguchi}}, \bibinfo {author} {\bibfnamefont
  {E.}~\bibnamefont {Kaxiras}},  \emph {et~al.},\ }\href@noop {} {\bibfield
  {journal} {\bibinfo  {journal} {Nature}\ }\textbf {\bibinfo {volume} {556}},\
  \bibinfo {pages} {80} (\bibinfo {year} {2018})}\BibitemShut {NoStop}%
\bibitem [{\citenamefont {Hopkins}\ \emph {et~al.}(1989)\citenamefont
  {Hopkins}, \citenamefont {Burns}, \citenamefont {Rimberg},\ and\
  \citenamefont {Westervelt}}]{Hopkins1989}%
  \BibitemOpen
  \bibfield  {author} {\bibinfo {author} {\bibfnamefont {P.~F.}\ \bibnamefont
  {Hopkins}}, \bibinfo {author} {\bibfnamefont {M.~J.}\ \bibnamefont {Burns}},
  \bibinfo {author} {\bibfnamefont {A.~J.}\ \bibnamefont {Rimberg}}, \ and\
  \bibinfo {author} {\bibfnamefont {R.~M.}\ \bibnamefont {Westervelt}},\
  }\href@noop {} {\bibfield  {journal} {\bibinfo  {journal} {Phys. Rev. B}\
  }\textbf {\bibinfo {volume} {39}},\ \bibinfo {pages} {12708} (\bibinfo {year}
  {1989})}\BibitemShut {NoStop}%
\bibitem [{\citenamefont {Koon}\ and\ \citenamefont
  {Castner}(1990)}]{Koon1990}%
  \BibitemOpen
  \bibfield  {author} {\bibinfo {author} {\bibfnamefont {D.~W.}\ \bibnamefont
  {Koon}}\ and\ \bibinfo {author} {\bibfnamefont {T.~G.}\ \bibnamefont
  {Castner}},\ }\href {\doibase 10.1103/PhysRevB.41.12054} {\bibfield
  {journal} {\bibinfo  {journal} {Phys. Rev. B}\ }\textbf {\bibinfo {volume}
  {41}},\ \bibinfo {pages} {12054} (\bibinfo {year} {1990})}\BibitemShut
  {NoStop}%
\bibitem [{\citenamefont {Friedman}\ and\ \citenamefont
  {Pollak}(1981)}]{Pollak1981}%
  \BibitemOpen
  \bibfield  {author} {\bibinfo {author} {\bibfnamefont {L.}~\bibnamefont
  {Friedman}}\ and\ \bibinfo {author} {\bibfnamefont {M.}~\bibnamefont
  {Pollak}},\ }\href {\doibase 10.1080/01418638108222584} {\bibfield  {journal}
  {\bibinfo  {journal} {Philosophical Magazine B}\ }\textbf {\bibinfo {volume}
  {44}},\ \bibinfo {pages} {487} (\bibinfo {year} {1981})},\ \Eprint
  {http://arxiv.org/abs/https://doi.org/10.1080/01418638108222584}
  {https://doi.org/10.1080/01418638108222584} \BibitemShut {NoStop}%
\bibitem [{\citenamefont {Galperin}\ \emph {et~al.}(1991)\citenamefont
  {Galperin}, \citenamefont {German},\ and\ \citenamefont
  {Karpov}}]{Galperin1991}%
  \BibitemOpen
  \bibfield  {author} {\bibinfo {author} {\bibfnamefont {Y.~M.}\ \bibnamefont
  {Galperin}}, \bibinfo {author} {\bibfnamefont {E.~P.}\ \bibnamefont
  {German}}, \ and\ \bibinfo {author} {\bibfnamefont {V.~G.}\ \bibnamefont
  {Karpov}},\ }\href@noop {} {\bibfield  {journal} {\bibinfo  {journal}
  {Zhurnal Eksperimentalnoi I Teoreticheskoi Fiziki}\ }\textbf {\bibinfo
  {volume} {99}},\ \bibinfo {pages} {343} (\bibinfo {year} {1991})}\BibitemShut
  {NoStop}%
\bibitem [{\citenamefont {{Lou}}\ and\ \citenamefont {{Xiang}}(2005)}]{CMom}%
  \BibitemOpen
  \bibfield  {author} {\bibinfo {author} {\bibfnamefont {P.}~\bibnamefont
  {{Lou}}}\ and\ \bibinfo {author} {\bibfnamefont {T.}~\bibnamefont
  {{Xiang}}},\ }\href@noop {} {\bibfield  {journal} {\bibinfo  {journal} {arXiv
  e-prints}\ ,\ \bibinfo {eid} {cond-mat/0501307}} (\bibinfo {year} {2005})},\
  \Eprint {http://arxiv.org/abs/cond-mat/0501307} {arXiv:cond-mat/0501307
  [cond-mat.mes-hall]} \BibitemShut {NoStop}%
\bibitem [{\citenamefont {Laughlin}(1981)}]{Laughlin1981}%
  \BibitemOpen
  \bibfield  {author} {\bibinfo {author} {\bibfnamefont {R.~B.}\ \bibnamefont
  {Laughlin}},\ }\href {\doibase 10.1103/PhysRevB.23.5632} {\bibfield
  {journal} {\bibinfo  {journal} {Phys. Rev. B}\ }\textbf {\bibinfo {volume}
  {23}},\ \bibinfo {pages} {5632} (\bibinfo {year} {1981})}\BibitemShut
  {NoStop}%
\bibitem [{\citenamefont {Von~Corbino}(1911)}]{Corbino}%
  \BibitemOpen
  \bibfield  {author} {\bibinfo {author} {\bibfnamefont {O.~M.}\ \bibnamefont
  {Von~Corbino}},\ }\href@noop {} {\bibfield  {journal} {\bibinfo  {journal}
  {Phys. Z.}\ }\textbf {\bibinfo {volume} {12}},\ \bibinfo {pages} {561}
  (\bibinfo {year} {1911})}\BibitemShut {NoStop}%
\bibitem [{\citenamefont {Perfetti}(2017)}]{PERFETTI2017171}%
  \BibitemOpen
  \bibfield  {author} {\bibinfo {author} {\bibfnamefont {M.}~\bibnamefont
  {Perfetti}},\ }\href {\doibase https://doi.org/10.1016/j.ccr.2017.08.013}
  {\bibfield  {journal} {\bibinfo  {journal} {Coordination Chemistry Reviews}\
  }\textbf {\bibinfo {volume} {348}},\ \bibinfo {pages} {171 } (\bibinfo {year}
  {2017})}\BibitemShut {NoStop}%
\bibitem [{\citenamefont {Chiaverini}\ \emph {et~al.}(2001)\citenamefont
  {Chiaverini}, \citenamefont {Yasumura},\ and\ \citenamefont
  {Kapitulnik}}]{PhysRevB.64.014516}%
  \BibitemOpen
  \bibfield  {author} {\bibinfo {author} {\bibfnamefont {J.}~\bibnamefont
  {Chiaverini}}, \bibinfo {author} {\bibfnamefont {K.}~\bibnamefont
  {Yasumura}}, \ and\ \bibinfo {author} {\bibfnamefont {A.}~\bibnamefont
  {Kapitulnik}},\ }\href {\doibase 10.1103/PhysRevB.64.014516} {\bibfield
  {journal} {\bibinfo  {journal} {Phys. Rev. B}\ }\textbf {\bibinfo {volume}
  {64}},\ \bibinfo {pages} {014516} (\bibinfo {year} {2001})}\BibitemShut
  {NoStop}%
\bibitem [{\citenamefont {Bleszynski-Jayich}\ \emph {et~al.}(2009)\citenamefont
  {Bleszynski-Jayich}, \citenamefont {Shanks}, \citenamefont {Peaudecerf},
  \citenamefont {Ginossar}, \citenamefont {von Oppen}, \citenamefont
  {Glazman},\ and\ \citenamefont {Harris}}]{Bleszynski-Jayich272}%
  \BibitemOpen
  \bibfield  {author} {\bibinfo {author} {\bibfnamefont {A.~C.}\ \bibnamefont
  {Bleszynski-Jayich}}, \bibinfo {author} {\bibfnamefont {W.~E.}\ \bibnamefont
  {Shanks}}, \bibinfo {author} {\bibfnamefont {B.}~\bibnamefont {Peaudecerf}},
  \bibinfo {author} {\bibfnamefont {E.}~\bibnamefont {Ginossar}}, \bibinfo
  {author} {\bibfnamefont {F.}~\bibnamefont {von Oppen}}, \bibinfo {author}
  {\bibfnamefont {L.}~\bibnamefont {Glazman}}, \ and\ \bibinfo {author}
  {\bibfnamefont {J.~G.~E.}\ \bibnamefont {Harris}},\ }\href {\doibase
  10.1126/science.1178139} {\bibfield  {journal} {\bibinfo  {journal}
  {Science}\ }\textbf {\bibinfo {volume} {326}},\ \bibinfo {pages} {272}
  (\bibinfo {year} {2009})},\ \Eprint
  {http://arxiv.org/abs/http://science.sciencemag.org/content/326/5950/272.full.pdf}
  {http://science.sciencemag.org/content/326/5950/272.full.pdf} \BibitemShut
  {NoStop}%
\bibitem [{\citenamefont {Finot}\ \emph {et~al.}(2008)\citenamefont {Finot},
  \citenamefont {Passian},\ and\ \citenamefont {Thundat}}]{s8053497}%
  \BibitemOpen
  \bibfield  {author} {\bibinfo {author} {\bibfnamefont {E.}~\bibnamefont
  {Finot}}, \bibinfo {author} {\bibfnamefont {A.}~\bibnamefont {Passian}}, \
  and\ \bibinfo {author} {\bibfnamefont {T.}~\bibnamefont {Thundat}},\ }\href
  {http://www.mdpi.com/1424-8220/8/5/3497} {\bibfield  {journal} {\bibinfo
  {journal} {Sensors}\ }\textbf {\bibinfo {volume} {8}},\ \bibinfo {pages}
  {3497} (\bibinfo {year} {2008})}\BibitemShut {NoStop}%
\bibitem [{\citenamefont {Breznay}\ \emph {et~al.}(2016)\citenamefont
  {Breznay}, \citenamefont {Steiner}, \citenamefont {Kivelson},\ and\
  \citenamefont {Kapitulnik}}]{Breznay2016}%
  \BibitemOpen
  \bibfield  {author} {\bibinfo {author} {\bibfnamefont {N.~P.}\ \bibnamefont
  {Breznay}}, \bibinfo {author} {\bibfnamefont {M.~A.}\ \bibnamefont
  {Steiner}}, \bibinfo {author} {\bibfnamefont {S.~A.}\ \bibnamefont
  {Kivelson}}, \ and\ \bibinfo {author} {\bibfnamefont {A.}~\bibnamefont
  {Kapitulnik}},\ }\href {\doibase 10.1073/pnas.1522435113} {\bibfield
  {journal} {\bibinfo  {journal} {PNAS}\ }\textbf {\bibinfo {volume} {113}},\
  \bibinfo {pages} {280} (\bibinfo {year} {2016})},\ \Eprint
  {http://arxiv.org/abs/https://www.pnas.org/content/113/2/280.full.pdf}
  {https://www.pnas.org/content/113/2/280.full.pdf} \BibitemShut {NoStop}%
\bibitem [{\citenamefont {Kurdesau}\ \emph {et~al.}(2006)\citenamefont
  {Kurdesau}, \citenamefont {Khripunov}, \citenamefont {Cunha}, \citenamefont
  {Kaelin},\ and\ \citenamefont
  {Tiwari}}]{kurdesau_khripunov_cunha_kaelin_tiwari_2006}%
  \BibitemOpen
  \bibfield  {author} {\bibinfo {author} {\bibfnamefont {F.}~\bibnamefont
  {Kurdesau}}, \bibinfo {author} {\bibfnamefont {G.}~\bibnamefont {Khripunov}},
  \bibinfo {author} {\bibfnamefont {A.~D.}\ \bibnamefont {Cunha}}, \bibinfo
  {author} {\bibfnamefont {M.}~\bibnamefont {Kaelin}}, \ and\ \bibinfo {author}
  {\bibfnamefont {A.}~\bibnamefont {Tiwari}},\ }\href {\doibase
  10.1016/j.jnoncrysol.2005.11.088} {\bibfield  {journal} {\bibinfo  {journal}
  {Journal of Non-Crystalline Solids}\ }\textbf {\bibinfo {volume} {352}},\
  \bibinfo {pages} {1466–1470} (\bibinfo {year} {2006})}\BibitemShut
  {NoStop}%
\end{thebibliography}%

\end{document}